\documentclass[referee,a4paper,fleqn,useAMS,usenatbib,onecolumn]{mnras}
\usepackage{graphicx}
\usepackage{psfrag}
\usepackage{bm}

\def\k{{\bf k}}

\def\kv{{\mathbf k}}

\def\U{{\bm{U}}}
\def\thetavec{{\bm{\theta}}}
\def\cl{C_{\ell}}

\def\n{\hat{\bm{n}}}

\def\V{\mathcal{V}}
\def\N{\mathcal{N}}
\def\u{{\bf U}} 

\newsavebox{\measurebox}
\begin{document}
\date {} 
\title[TGE] {A Tapered Gridded Estimator (TGE) for the Multi-Frequency Angular Power Spectrum (MAPS) and the Cosmological 
HI 21-cm Power Spectrum} 
\author[S. Bharadwaj et al.]{Somnath Bharadwaj$^{1}$\thanks{Email:somnath@phy.iitkgp.ac.in}, Srijita Pal$^{1}$, Samir Choudhuri$^{2}$ and  Prasun Dutta$^{3}$\\
$^{1}$ Department of Physics,  \& Centre for Theoretical Studies, IIT Kharagpur,  Kharagpur 721 302, India\\
$^{2}$ National Centre For Radio Astrophysics, Post Bag 3, Ganeshkhind, Pune 411007, India\\
$^{3}$ Department of Physics, IIT (BHU), Varanasi 221005, India\\}

\maketitle

\begin{abstract}
In this work we present a new approach to estimate the power spectrum $P(\k)$ of redshifted HI 21-cm brightness temperature fluctuations. 
The MAPS  $\cl(\nu_a,\nu_b)$  completely quantifies the second order statistics of the sky signal under the assumption that the signal is statistically homogeneous and isotropic on the sky. Here we generalize an already existing visibility based estimator for $\cl$, namely TGE, to develop an estimator for   $\cl(\nu_a,\nu_b)$ .
The 21-cm power spectrum  is the Fourier transform of $\cl(\Delta \nu)$ with respect to 
 $\Delta \nu=\mid \nu_a-\nu_b \mid$, and we use this to estimate $P(\k)$. Using simulations 
 of $150 \, {\rm MHz}$ GMRT observations, we find that this estimator is able to recover $P(k)$ 
 with an accuracy of $5-20 \%$ over a reasonably large $k$ range even when the data in $80 \%$ randomly chosen frequency channels is flagged. 
\end{abstract} 


\section{Introduction}
\label{intro}
Measurements of the cosmological HI 21-cm  power spectrum can be used to probe the large scale distribution of neutral hydrogen (HI) 
across a large redshift range from the Dark Ages to the Post-Reionization Era
(e.g. \citealt{BA5,furla06,morales10,prichard12,mellema13}). Being very faint in nature, the 21-cm signal is buried in foregrounds which are four to five orders of magnitude larger than the expected signal  (\citealt{shaver99,santos05,ali,bernardi09,ghosh150,iacobelli13,samir17a}). There are several ongoing and future experiments, e.g. Donald C. Backer Precision Array to Probe
the Epoch of Reionization (PAPER{\footnote{http://astro.berkeley.edu/dbacker/eor}},
\citealt{parsons10}), the Low Frequency Array
(LOFAR{\footnote{http://www.lofar.org/}}, \citealt{haarlem,yata13}),
the Murchison Wide-field Array
(MWA{\footnote{http://www.mwatelescope.org}},
\citealt{bowman13,tingay13}), the Giant Metrewave Radio Telescope
(GMRT, \citealt{swarup})  the
Square Kilometer Array (SKA1
LOW{\footnote{http://www.skatelescope.org/}}, \citealt{koopmans15})
and the Hydrogen Epoch of Reionization Array
(HERA{\footnote{http://reionization.org/}}, \citealt{deboer17}) which are 
aiming to detect the 21-cm power spectrum from the Epoch of Reionization (EoR).

The biggest challenge for a detection of the redshifted 21-cm signal are the foregrounds which include point sources, the diffuse Galactic synchrotron emission, the free-free emission from our Galaxy and external galaxies. Various techniques have been proposed to overcome this issue.  The  foreground subtraction technique proposes to subtract a foreground model from the visibility data or the image and use the residual data to detect the 21-cm power spectrum  \citep{jelic08,bowman09,paciga11,chapman12,trott1,paciga13,trott16}. 
Considering $P(k_{\perp},k_{\parallel})$, the  cylindrical power spectrum  of the 21-cm brightness temperature fluctuations, the foregrounds are expected to be primarily confined to a wedge in the $(k_{\perp},k_{\parallel})$ plane. Here, $k_{\perp}$ ans $k_{\parallel}$ refer to the components of the 3-dimensional wave vector ${\bf k}$ perpendicular and parallel to the line of sight direction respectively. The foreground avoidance technique proposes to 
use the region outside this ``Foreground Wedge" to estimate the 21-cm power spectrum  
 \citep{adatta10,parsons12,vedantham12,pober13,thyag13,parsons14,pober14,liu14a,liu14b,dillon14,dillon15,zali15}.

A large variety of  estimators have been proposed and applied to measure the power spectrum of the brightness temperature fluctuations 
using the visibility data measured in radio interferometric observations. Image-based estimators \citep{seljak97,paciga13} have the deconvolution error which arises during image reconstruction, and this may affect the estimated power spectrum. There are a few other techniques, like the Optimal Mapmaking Formalism \citep{morales09} where the deconvolution errors can be avoided during imaging. It is possible to overcome this issue by estimating the power spectrum directly from  the measured visibilities \citep{morales05,mcquinn06,pen09,liu12,parsons12,liu14a,liu14b,dillon15,trott16}. 
\citet{liu16} have proposed  an estimator which uses the spherical Fourier-Bessel basis to account for  sky curvature. 
In addition to the sky signal, the visibilities (or the image) also have a noise contribution, and  the noise bias is an important 
issue for power spectrum estimation. For example, \citet{zali15} have divided the data sets into even and odd LST bins and have  correlated these  to avoid introducing a noise bias. This approach however does not utilize the full signal available in the data. 
The foreground contributions from the outer regions  of the telescope's field of view  (including the side-lobes) pose a severe 
problem for detecting the cosmological 21-cm signal \citep{pober16}. 
In this paper we develop on the visibility based Tapered Gridded Estimator (TGE; \citealt{samir14},\citealt{samir16b}, hereafter Papers I and II respectively) whose salient features we summarize as follows. First,  it uses the data to internally estimate the noise bias and subtracts this out to provide an unbiased estimate of the power spectrum. Second, it deals with the gridded visibilities which makes it computationally efficient. Third, it 
tapers the sky response to suppress the contribution from the outer regions of the telescope's field of view.

Nearly all the estimators for $P(k_{\perp},k_{\parallel})$, including the 3D TGE  (Paper II), consider a Fourier transform of the measured visibilities $\V(\U,\nu)$ along the frequency axis $\nu$ to obtain the visibilities $\V(\U,\tau)$  in delay space $\tau$ \citep{morales04}. This  is used to estimate $P(k_{\perp},k_{\parallel})$. A difficulty arises if the data is missing or flagged in a few frequency  channels in which case the  delay channel visibilities $\V(\U,\tau)$  and the estimated power spectrum 
$P(k_{\perp},k_{\parallel})$ are both modified by a convolution with the Fourier transform of the  frequency sampling function. Missing or flagged channels are quite common in any typical observation due to a variety of reasons including man made radio frequency interference (RFI). 
The CHIPS estimator developed by \citet{trott16} overcomes this problem by using  Least-Squares Spectral Analysis (LSSA)   to evaluate $\V(\U,\tau)$. However this needs to be applied  individually for each baseline, and  the entire process could be computationally expensive for large data volumes. In this paper we propose an alternative approach to estimate $P(k_{\perp},k_{\parallel})$ which is able to handle the problem of missing or flagged data with relative ease. Another point to note is that the earlier estimators all introduce a frequency filter which smoothly goes to zero at the two edges of the frequency band. This is introduced to avoid a discontinuity at the edges of the band, however it results in the loss of some signal. Such a filter is not needed in the new estimator proposed here.  

The multi-frequency angular power spectrum $\cl(\nu_a,\nu_b)$ (MAPS; \citealt{maps},\citealt{mondal2018}) completely quantifies the second order statistics of the sky signal under the assumption that the signal is statistically homogeneous and isotropic on the sky. This however does not assume that the signal is ergodic or statistically homogeneous along the frequency axis. We have $\cl(\nu_a,\nu_b) = \cl(\Delta \nu)$ where $\Delta \nu=\mid \nu_a-\nu_b \mid$ if we impose the additional condition that the signal is ergodic along frequency. The 3D 21-cm power spectrum $P(k_{\perp},k_{\parallel})$ is the Fourier transform of $\cl(\Delta \nu)$. In the new approach presented here we first estimate $\cl(\Delta \nu)$ and use the binned  $\cl(\Delta \nu)$ to estimate $P(k_{\perp},k_{\parallel})$. 
Even if some channels are missing, it is quite possible that the frequency separations $\Delta \nu$ are all present in the data. In this case it is quite straight forward to evaluate $P(k_{\perp},k_{\parallel})$ through a Fourier transform of $\cl(\Delta \nu)$.
More sophisticated techniques like the LSSA can be used in case some $\Delta \nu$ are missing, however this needs to be applied to the binned $\cl(\Delta \nu)$ and the task is not computationally expensive. 

The MAPS  $\cl(\Delta \nu)$ has been used  to quantify the statistical properties of the background radiation in GMRT observations at 150  MHz \citep{ali,ghosh150} and 610 MHz \citep{ghosh1,ghosh2}. The HI signal contribution to the measured $\cl(\Delta \nu)$ is expected to decorrelate rapidly when $\Delta \nu$ is increased whereas the foreground contribution is expected to remain correlated for large $\Delta \nu$ separations. This property was used \citep{ghosh2} to model and remove the foreground contribution and obtain a residual $\cl(\Delta \nu)$ which is consistent with noise. It was thereby possible to place an observational limit on the HI 21-cm power spectrum at $z \approx 1.3$. The estimator used in these earlier works individually correlates pairs of visibilities to estimate $\cl(\Delta \nu)$, a technique which is computationally expensive. The 2D TGE (Paper II) presents an efficient technique to estimate the angular power spectrum $\cl$.  
In Section 2. of this paper we have generalized this earlier work to develop an estimator for the MAPS $\cl(\nu_a,\nu_b)$.  In Section 3. we present how $P(k_{\perp},k_{\parallel})$ is obtained from the estimated $\cl(\Delta \nu)$. Section 4. presents the Simulations which we have used to validate our estimator, Section 5. presents the Results and Section 6. presents the Discussion and Conclusions.

We have used the cosmological parameters from the (Planck +
WMAP) best-fit $\Lambda$CDM cosmology (\citealt{ade15}) throughout this paper.

\section{An overview of the Tapered Gridded Estimator}
\label{overview}

The 2D TGE, presented  in Paper II considers radio-interferometric observations
at a single frequency $\nu$ and   uses the measured visibilities $\V_i$
to estimate the angular power spectrum $\cl$ of the background  radiation at the frequency $\nu$. 
Here $\V_i$ refers to the $i$-th visibility measurement with a corresponding 
baseline $\u_i$. The measured visibilities can
be expressed as

\begin{equation}
\V_i= \left( \frac{\partial B}{\partial T} \right) \int \, d^2 U \,
\tilde{a}\left(\u_i - \u\right)\, \, \Delta \tilde{T}(\u)+\N_i.
\label{eq:1}
\end{equation}
Here, the first term is the sky signal which is the convolution of 
$\tilde{a}\,(\u)$ and $\Delta \tilde{ T}(\u)$ where these are the Fourier transforms of the primary beam ${\cal A}(\thetavec)$ and the temperature fluctuations in the sky $\delta T(\thetavec)$ respectively, and $B=2k_B T/\lambda^2$ is the Planck function in
the Rayleigh-Jeans limit. The second term $\N_i$ is the            system  noise contribution.

In order to taper the sky response,  
the measured visibilities are convolved with a function 
$\tilde{w}(\u)$ which is the Fourier transform of a window function 
${\cal W}(\thetavec)$  which falls off to a value close to zero well before the first null of the telescope's 
primary beam pattern (Paper I). Further, in order to reduce the computation,
the convolved visibilities are evaluated on a grid in $uv$ space
using   
\begin{equation}
\V_{cg} = \sum_{i}\tilde{w}(\u_g-\u_i) \, \V_i \,.
\label{eq:a1}
\end{equation}
where the `c' in $\V_{cg}$ refers to ``convolved'' and $g$ refers to different  grid points with corresponding baselines
$\u_g$. The sky response of $\V_{cg}$ is tapered with the window function
${\cal W}(\thetavec)$. Here we have used ${\cal W}(\thetavec)=e^{-\theta^2/\theta_w^2}$
where the value of $\theta_w=57^{'}$ is chosen so as to 
suppress the contribution from the outer regions and sidelobes of the 
telescope's primary beam pattern (Figure 1 of \citealt{samir16a}). For comparison, the full width 
half maxima of the $150 \, {\rm MHz}$ GMRT primary beam pattern may be estimated to be $1.03 \lambda/D=157^{'}$ 
where $D=45 \, {\rm m}$ is the antenna diameter.

The convolved gridded visibilities can be expressed as
\begin{equation}
\V_{cg}= \left( \frac{\partial B}{\partial T} \right)  \int \, d^2 U  \, 
 \tilde{K}\left(\u_g - \u\right)\,  \, \Delta \tilde{T}(\u) + \sum_{i}\tilde{w}(\u_g-\u_i) \, \N_i,  
\label{eq:6}
\end{equation}
where 
\begin{equation}
\tilde{K}\left(\u_g - \u\right)= \int d^2 U^{'} 
\tilde{w}(\u_g-\u^{'}) B(\u^{'}) \tilde{a}\left(\u^{'} - \u\right) 
\label{eq:7}
\end{equation}
is an effective  ``gridding kernel'', and 
\begin{equation}
{\rm B}(\u)=\sum_i \delta^2_D(\u-\u_i)
\label{eq:8}
\end{equation}
 is the  baseline sampling function of the measured visibilities.

The 2D TGE estimator is defined as 
\begin{equation}
{\hat E}_g= M_g^{-1} \, \left( \mid \V_{cg} \mid^2 -\sum_i \mid
\tilde{w}(\u_g-\u_i) \mid^2  \mid \V_i \mid^2 \right) \,. 
\label{eq:a2}
\end{equation}
with $\langle {\hat E}_g \rangle = \cl{_g}$ where 
$\ell_g=2 \pi U_g$, and $\langle \,\rangle$ denotes an ensemble average 
over multiple realizations of the sky brightness temperature 
fluctuations which are recorded in the visibilities. 
The second term in the brackets $(...)$ in eq.~(\ref{eq:a2}) 
is introduced 
to subtract out the noise bias contribution which arises due to the 
correlation of a visibility with itself. $M_g$ is a normalization factor 
which we shall discuss later. Simulations  show 
that the 2D TGE provides an unbiased estimate of the angular power 
spectrum $\cl$ (Paper II)
while effectively suppressing the contribution from the 
sidelobes and outer regions of the telescope's primary beam \citep{samir17b}. 

\subsection{$M_g$ Calculation}

As discussed in Paper II, the normalization constant $M_g$ can be written as,
\begin{equation}
M_g=V_{1g} - \sum_i \mid \tilde{w}(\u_g-\u_i) \mid^2 V_0
\label{eq:12}
\end{equation}
where, 
\begin{equation}
V_{1g}= \left( \frac{\partial B}{\partial T} \right)^2
 \int d^2 U \, \mid \tilde{K}(\u_i-\u) \mid^2 \,.
\label{eq:10}
\end{equation}
and 
\begin{equation}
V_0= \left( \frac{\partial B}{\partial T} \right)^2 \int d^2 U \, \mid
\tilde{a}(\u_i-\u) \mid^2 \,.
\label{eq:3}
\end{equation}
The values of $M_g$ (eq. \ref{eq:12}) depend on the baseline
distribution (eq. \ref{eq:8}) and the form of the tapering function
${\cal W}(\theta)$, and it is necessary to calculate $M_g$ at every
grid point in the $uv$ plane.  Paper I presents an
analytic approximation to  estimate $M_g$. 
While this has been found to work very well in a situation where the
baselines have a nearly uniform and dense $uv$ coverage (Fig. 7 of
Paper I), it leads to $C_{\ell}$  being  overestimated in a situation 
where we have a
sparse and non-uniform $uv$ coverage. Paper II presents a different
method to estimate $M_g$ which has been found to  work well even
if the $uv$ coverage is  sparse and non-uniform .

We now briefly present how the normalization constant $M_g$ is calculated
for $\cl$ estimation in eq.~(\ref{eq:a2}) . As discussed in Paper II, 
we proceed by constructing random  realizations  of 
simulated  visibilities $[\V_i]_{\rm UAPS}$
 corresponding to a situation where the sky signal has 
  an unit angular power spectrum  (UAPS) 
 $\cl=1$.  The simulated visibilities have   
 exactly  the same baseline distribution as the actual observed
  visibilities. 
We then have (eq. \ref{eq:a2}) 
\begin{equation}
M_g=\langle \left( \mid \V_{cg} \mid^2 -\sum_i \mid \tilde{w}(\u_g-\u_i) \mid^2 \langle \mid \V_i \mid^2 \right)  \rangle_{\rm UPAS} 
\label{eq:m1}
\end{equation}
which allows us to estimate $M_g$. We average over $N_u$ independent
realizations of the UPAS to reduce the statistical uncertainty.

\subsection{Binning}
The estimator ${\hat E}_g$ provides an estimate of $\cl$ at different grid points $\u_g$ on the $uv$ plane. We have binned 
the estimates in order to increase the signal to noise ratio and also reduce the data volume.  The signal is assumed to 
be statistically isotropic on the sky   whereby it is independent of the direction of $\u_g$.  This allows us to average 
the $\cl$ estimates within an annular region on the $uv$ plane.    We  define the binned Tapered Gridded Estimator for bin $a$ 
using 
\begin{equation}
{\hat E}_G(a) = \frac{\sum_g w_g  {\hat E}_g}{\sum_g w_g } \,.
\label{eq:a15}
\end{equation}
where $w_g$ refers to the weight assigned to the contribution from any particular 
grid point.  The choice $w_g=1$ assigns equal weightage to the value of $C_{\ell_g}$
estimated at each grid point, whereas $w_g=M_g$ corresponds to a situation where the grid points 
which have a denser baseline sampling (less system noise) would be given a larger weightage. The former would be desireable
if one wishes to optimize with respect to the cosmic variance whereas the latter would be preferred to optimize 
with respect to the system noise contribution.  The optimum choice of $w_g$ to maximize the signal to noise ratio would 
depend on the window function and the  baseline distribution, and we plan to address this in future.  

The binned estimator  has an expectation value 
\begin{equation}
\bar{C}_{\bar{\ell}_a}  = \frac{ \sum_g w_g C_{\ell_g}}{ \sum_g w_g}
\label{eq:ga16}
\end{equation}
where $ \bar{C}_{\bar{\ell}_a}$ is the average  angular power spectrum  at 
 \begin{equation}
\bar{\ell}_a =
\frac{ \sum_g w_g \ell_g}{ \sum_g w_g}
\label{eq:a18}
\end{equation}
which is the   effective angular multipole  for bin $a$. 

\section{The multi-frequency angular power spectrum}
\label{maps}
The multi-frequency angular power spectrum $\cl(\nu_a,\nu_b)$ 
\citep{maps} characterizes the joint frequency and angular dependence 
of the statistical properties of the 
background sky signal. 
We decompose the brightness temperature fluctuations
$\delta T_{\rm b} (\n,\,\nu)$ in terms of spherical harmonics
$Y_{\ell}^{\rm m}(\n)$ using 
\begin{equation}
\delta T_{\rm b} (\n,\,\nu)=\sum_{\ell,m} a_{\ell {\rm m}} (\nu) \,
Y_{\ell}^{\rm m}(\n)
\label{eq:alm}
\end{equation}
and define the multi-frequency angular power spectrum (hereafter MAPS) as 
\begin{equation}
\cl(\nu_a, \nu_b) = \big\langle a_{\ell {\rm m}} (\nu_a)\, a^*_{\ell
  {\rm m}} (\nu_b) \big\rangle\, .
\label{eq:cl}
\end{equation}
As discussed in \citet{mondal2018}, 
we expect  $\cl(\nu_1,\nu_2)$ to entirely quantify the second order statistics 
of the redshifted  21-cm signal.

We  now proceed to define  a 
visibility based Tapered Gridded Estimator (TGE) for $\cl(\nu_a,\nu_b)$.
We  generalize the analysis to consider visibility 
measurements $\V_i(\nu_a)$ at multiple frequency channels  
$1 \le a \le N_c$, each of  width $\Delta \nu_c$, with $N_c$ channels
that span a bandwidth $B_{bw}$.  Here we allow for the possibility that 
several of the data are bad or missing. We assume that such data has
been identified and flagged, and this information is stored using a
flagging variable $F_i(\nu_a)$ which has value $0$ for the flagged data 
and value $1$ otherwise. We then have 
\begin{equation}
\V_{cg}(\nu_a) = \sum_{i}\tilde{w}(\u_g-\u_i) \, \V_i(\nu_a) \,F_i(\nu_a).
\label{eq:a3}
\end{equation}
which  allows us to define the Tapered Gridded Estimator (TGE) for 
 $\cl(\nu_a,\nu_b)$ as 
\begin{equation}
{\hat E}_g(\nu_a,\nu_b)= M_g^{-1}(\nu_a,\nu_b) \, 
{\mathcal Re} \left(  \V_{cg}(\nu_a) \,  \V_{cg}^{*}(\nu_b)
-\delta_{a,b} \, \sum_i F_i(\nu_a)  \mid
\tilde{w}(\u_g-\u_i) \mid^2  
  \mid  \V_i(\nu_a)  \mid^2 \right) \,.  
\label{eq:a4}
\end{equation}
where ${\mathcal Re}()$ denotes the real part, $\delta_{a,b}$ is a Kronecker delta {\it i.e.} it is necessary to 
subtract the noise bias only when the two frequencies are the same 
$(\nu_a=\nu_b)$, and the noise
in the visibility measurements at two different frequencies 
$(\nu_a \neq \nu_b)$  are uncorrelated. 

The TGE defined in eq.~(\ref{eq:a4}) 
provides an unbiased estimate of $\cl{_g}(\nu_a,\nu_b)$ at the angular
 multipole $\ell_g=2 \pi U_g$ {\it i.e.} 
\begin{equation}
\langle {\hat E}_g(\nu_a,\nu_b) \rangle = \cl{_g}(\nu_a,\nu_b)
\label{eq:a5}
\end{equation}
 We  use this to define the binned Tapered Gridded Estimator for bin $a$ 
\begin{equation}
{\hat E}_G[a](\nu_a,\nu_b) = \frac{\sum_g w_g  {\hat E}_g(\nu_a,\nu_b)}
{\sum_g w_g } \,.
\label{eq:a6}
\end{equation}
where $w_g$ refers to the weight assigned to the contribution from any particular 
grid point $g$. For the analysis presented in this paper we have used the 
weight $M_g(\nu_a,\nu_b)$  which roughly averages the visibility correlation 
$\V_{cg}(\nu_a) \,  \V_{cg}^{*}(\nu_b)$ across  all the  grid points
 which are sampled by the baseline distribution. 
The binned estimator  has an expectation value 
\begin{equation}
\bar{C}_{\bar{\ell}_a}(\nu_a,\nu_b)
  = \frac{ \sum_g w_g \cl{_g}(\nu_a,\nu_b)}{ \sum_g w_g}
\label{eq:a7}
\end{equation}
where $ \bar{C}_{\bar{\ell}_a}(\nu_a,\nu_b)$ is the bin averaged
  multi-frequency angular  power spectrum  (MAPS) at 
 \begin{equation}
\bar{\ell}_a =
\frac{ \sum_g w_g \ell_g}{ \sum_g w_g}
\label{eq:a8}
\end{equation}
which is the   effective angular multipole  for bin $a$. 

Paper II describes how we have estimated $M_g$ using UAPS simulations in the context of observations
at a single frequency. This has also  been summarized in Section \ref{overview} of this paper. 
Here we have extended the earlier analysis to simulate visibilities 
$[\V_i(\nu_a)]_{\rm UMAPS}$ for which we have an unit multi-frequency
angular power spectrum $\cl(\nu_a,\nu_b)=1$. We also apply the same
flagging variable $F_i(\nu_a)$ as the actual data to the simulated
data. Using the simulated  visibilities $[\V_i(\nu_a)]_{\rm UMAPS}$ and the
actual flagging variable $F_i(\nu_a)$ in eq.~(\ref{eq:a4}),
we have an estimate of $M_g(\nu_a,\nu_b)$.
 We have used multiple realizations of the simulations
to reduce the uncertainty in the estimated values of $M_g(\nu_a,\nu_b)$.

We note that the estimator presented  here  does not take into account the fact that the baselines $\u_i=\bf{d_i}/\lambda$ (where $\bf{d}$ is the antenna spacing) and the primary beam pattern ${\cal A}(\thetavec,\nu)$ both change with frequency and these
are held fixed at the values corresponding to the central frequency $\nu_c$. While this may not have a very significant effect on the recovered 21-cm power spectrum, it is very important for the foregrounds where this leads to the foreground wedge 
(eg. \citealt{adatta10,parsons12,vedantham12}).  
We note that the frequency dependence of the baselines has been included  in earlier versions of the MAPS estimator  \citep{ali,ghosh1,ghosh150} which did not incorporate gridding and tapering.                                
It is possible to incorporate the frequency dependence of  the baselines in the TGE  by suitably scaling the baselines $\u_i$ at the time of convolution and gridding (eq. \ref{eq:a3}), and we plan to address this in future work.

\section{Estimating $P(k_{\perp},\,k_{\parallel})$}
In order to estimate the 3D power spectrum $P(k_{\perp},\,k_{\parallel})$
we assume that the redshifted 21-cm signal  is statistically
homogeneous (ergodic) along the line of sight (e.g. \citealt{mondal2018}). We then have
$\cl(\nu_a,\nu_b)=\cl(\Delta \nu)$ where $\Delta \nu=  \mid \nu_b-\nu_a\mid$
{\it i.e.} the statistical properties of the signal depends only
on the  frequency separation and not the individual frequencies. 
In the flat sky approximation,  the power spectrum $P(k_{\perp},\,k_{\parallel})$ 
of the brightness temperature fluctuations of the redshifted 21-cm   signal 
is  the Fourier transform of
$\cl(\Delta \nu)$, and we have \citep{maps}  
\begin{equation}
P(k_{\perp},\,k_{\parallel})= r^2\,r^{\prime} \int_{-\infty}^{\infty}  d (\Delta \nu) \,
  e^{-i  k_{\parallel} r^{\prime} \Delta  \nu}\, \cl(\Delta \nu)
\label{eq:cl_Pk}
\end{equation}
where $k_{\parallel}$ and $k_{\perp}=\ell/r$ are the
components of $\kv$ respectively parallel and perpendicular to the
line of sight, $r$ is the comoving distance corresponding to $\nu_c$ 
the central  
frequency of our observations and $r^{\prime}~(=d r/d \nu)$ is evaluated at $\nu_c$. 
A brief derivation of eq.~(\ref{eq:cl_Pk}) is also presented in the
Appendix of \citet{mondal2018}. In this paper we have used
(eq.~\ref{eq:cl_Pk}) to  estimate 
$P(k_{\perp},\,k_{\parallel})$ from  the MAPS $\cl(\nu_a,\nu_b)$.

First we impose the ergodic assumption on
$\cl(\nu_a,\nu_b)$ which has been estimated from the visibility data 
using eq.~(\ref{eq:a4}) and binned 
using eq.~(\ref{eq:a6},\ref{eq:a7} and \ref{eq:a8}).
For a fixed $\ell$ and $\Delta \nu$, we average over all the 
$\cl(\nu_a,\nu_b)$ values  for which $\mid \nu_b-\nu_a \mid=\Delta \nu$
to obtain $\cl(\Delta \nu)$. We then have
$\cl(n \, \Delta \nu_c)$ where $-(N_c-1) \le n \le (N_c-1)$ with
$\cl(n \, \Delta \nu_c) =\cl(-n \, \Delta \nu_c)$. We see that 
$\cl(n \, \Delta \nu_c)$ is a periodic function of $n$ 
with period $2N_c-2$.   We use the discrete Fourier transform 
\begin{equation}
\bar{P}(k_{\perp},\,k_{\parallel m}) = (r^2\,r^{\prime} \, \Delta \nu_c) 
\sum_{n=-N_c+2}^{N_c-1}   \exp \left( -i  k_{\parallel m} r^{\prime} \, n \Delta  
\nu_c \right) \, \cl(n \Delta \nu_c) 
\label{eq:b1}
\end{equation}
with $k_{\parallel m}=m \times [\pi/r^{\prime}_{\rm c} \, \Delta \nu_c (N_c-1)]$
to estimate $\bar{P}(k_{\perp},\,k_{\parallel m})$ which  is already binned in 
$k_{\perp}$. We have further binned in $k_{\parallel m}$ to obtain the 
Spherical Power Spectrum $P(k)$, and the Cylindrical Power Spectrum
 $P(k_{\perp},\,k_{\parallel})$. 

\section{Simulations}
We have carried out simulation to validate the estimator presented here. 
We have simulated $8$ hours of $150 \, {\rm MHz}$ Giant Meterwave Radio 
Telescope \citep{swarup} observations with $N_c=257$ channels of width
$\Delta \nu_c=62.5 \, {\rm KHz}$ spanning $B_{bw} \approx 16 \, {\rm MHz}$ 
and integration time $\Delta t =16 \, {\rm s}$ towards 
RA=10h46m00s and DEC=59$^{\circ}$00$^{'}$59$^{''}$. We note that the EoR 21-cm signal is not expected to be ergodic over the $16 \, {\rm MHz}$ bandwidth considered here due to the Light Cone effect \citep{mondal2018}. 
However, we have not considered this effect here and assumed that the signal is ergodic. The sky signal, we assume, 
is entirely the redshifted HI 21-cm emission whose brightness temperature
fluctuations are characterized by the 3D power spectrum $P^m(k)=(k/k_0)^{n} \,
{\rm mK}^2 \, {\rm Mpc}^3$. For the purpose of this paper we have arbitrarily
chosen the values $k_0=(1.1)^{-1/2} \, {\rm Mpc}^{-1}$ and $n=-2$.  We have followed the 
procedure outlined in Section 4 of \citet{samir16b} to simulate
visibilities $\V_i(\nu_a)$ corresponding to different statistically 
independent realizations of the brightness temperature fluctuations. 

In addition to the sky signal, the visibilities also contain a system 
noise contribution. We have modelled the system noise contribution 
to the visibilities  as  Gaussian random 
variables whose real and imaginary parts both have zero  mean and variance
$\sigma^2_N$. For comparison we have also estimated $\sigma_{sky}^2$ which 
is the same quantity  for the simulated sky signal contribution.
The ratio $R=\sigma_N/\sigma_{sky}$ gives an estimate of the relative
contribution of the system noise with respect to the sky signal. In our
simulations we have used $R=10$ which corresponds to a situation where 
the noise contribution to an individual visibility is $R=10$ times the 
sky signal contribution. We have generated  $24$  
statistically independent realizations of both the sky signal and 
the system noise. The resulting $24$ statistically independent 
realizations of the  simulated visibilities were used to estimate
the mean and  $1-\sigma$ errors  for the results presented below. 
We have considered simulations both with and without flagging. 
For each baseline we have generated 
random integers in the range $1 \le a \le N_c$ and flagged the corresponding 
channels.  We have carried out simulations for various values of $f_{\rm FLAG}$ 
(the fraction of flagged channels) in the range $0 \le f_{\rm FLAG} \le 0.8$.

We note that the frequency dependence of the baselines $\u=\bf{d}/\lambda$ and the 
primary beam pattern ${\cal A}(\thetavec,\nu)$ have both been incorporated in the simulated 
visibilities.

\begin{figure*}
\begin{center}
\psfrag{freq. seper.}[1.0]{$\Delta\nu\hspace{.2cm} \textrm{MHz}$}
\psfrag{l=2.165176e+03}[c][1.0]{$\ell=2165$}
\psfrag{l=5.255096e+03}[c][1.0]{$\ell=5255$}
\psfrag{label}[c][1.0]{\hspace{2cm}$\textrm{Noise, 80\% Flagging}$}
\psfrag{Noise, 80
\psfrag{-5e-09}[c][1.0]{$-0.5\hspace{.5cm}$}\psfrag{ 0}[r][1.0]{$ 0.0\hspace{.1cm}$}\psfrag{ 5e-09}[c][1.0]{$ 0.5\hspace{.3cm}$}\psfrag{ 1e-08}[c][1.0]{$ 1.0\hspace{.3cm}$}\psfrag{ 1.5e-08}[c][1.0]{$ 1.5\hspace{.3cm}$}\psfrag{ 2e-08}[c][1.0]{$ 2.0\hspace{.3cm}$}\psfrag{ 2.5e-08}[c][1.0]{$ 2.5\hspace{.3cm}$}\psfrag{ 3e-08}[c][1.0]{$ 3.0\hspace{.3cm}$}
\psfrag{Cl}[1.0]{$\cl(\Delta\nu)\hspace{.1cm}10^{-8}\hspace{.1cm}\textrm{mK}^2$}
\psfrag{freq. seper.}[c][1.0]{$\Delta\nu\hspace{.2cm}\textrm{MHz}$}
\psfrag{l}[1.0]{$\ell$}
\psfrag{ 0.01}[c][1.0]{$ 0.01$}\psfrag{ 0.1}[c][1.0]{$ 0.1$}\psfrag{ 1}[c][1.0]{$ 1$}\psfrag{ 10}[c][1.0]{$ 10$}
\psfrag{10}{$ $}
\psfrag{2}[1.0]{$10^{2}$}\psfrag{3}[1.0]{$10^{3}$}\psfrag{4}[1.0]{$10^{4}$}
\includegraphics[scale=0.8,angle=0]{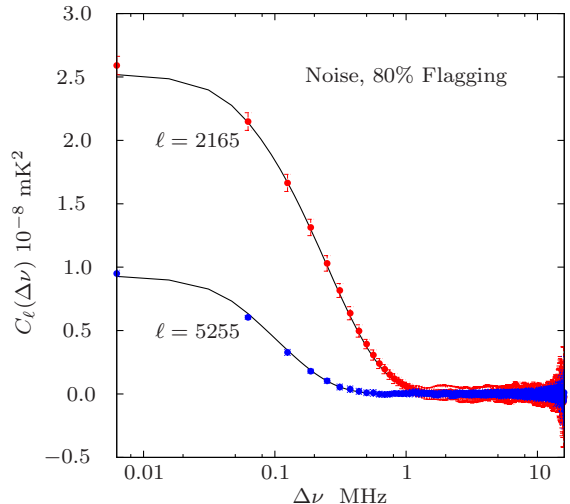}
\caption{This shows $\cl(\Delta\nu)$ as a function of $\Delta\nu$ for two values of $\ell$.
The data points with $1-\sigma$ error-bars are estimated from 24 realizations of the simulations.
 Note that the $\Delta \nu=0$ points have been slightly shifted for convenience of plotting on a logarithmic scale. 
 The lines show the theoretical predictions calculated by using the input model power spectrum 
 $P^m(k)$ in eq.~(\ref{eq:c1}).}
\label{fig:1}
\end{center}
\end{figure*}

\begin{figure*}
\begin{center}
\psfrag{Cl}[1.0]{$\cl(\Delta\nu)\hspace{.1cm}10^{-8}\hspace{.1cm}\textrm{mK}^2$}
\psfrag{freq. seper.}[1.0]{$\Delta\nu\hspace{.2cm} \textrm{MHz}$}
	\psfrag{ 0}[l][1.0]{$\hspace{.1cm} 0$}
	\psfrag{-5}[l][1.0]{$ $}\psfrag{ 0}[c][1.0]{$ 0$}\psfrag{ 5}[l][1.0]{$ $}\psfrag{ 10}[c][1.0]{$ 10$}\psfrag{ 15}[l][1.0]{$ $}\psfrag{ 20}[c][1.0]{$ 20$}\psfrag{ 25}[l][1.0]{$ $}\psfrag{ 30}[c][1.0]{$ 30$}\psfrag{ 35}[l][1.0]{$ $}\psfrag{ 40}[c][1.0]{$ 40$}\psfrag{ 45}[l][1.0]{$ $}
	\psfrag{ 0.01}[c][1.0]{$ 0.01$}
	\psfrag{l}[1.0]{$\ell$}
	\includegraphics[scale=1.0,angle=0]{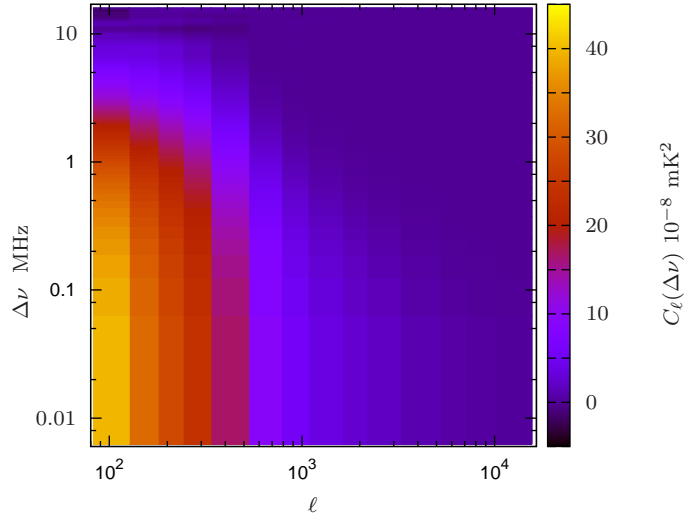}
\caption{This shows $\cl(\Delta\nu)$ across the entire $\ell$ and $\Delta \nu$ range considered here for simulations with noise and $80 \%$ flagging.  We have shifted the $\Delta\nu=0$ values to $\Delta\nu=0.006251$\textrm{MHz} for convenience of plotting on a logarithmic scale.}
\label{fig:2}
\end{center}
\end{figure*}

\section{Results}
The analysis here was restricted to baselines in the range $10 \le \mid \u_i \mid \le 3,000$, and the $uv$ plane was divided into $15$ annular bins at equal logarithmic intervals for power spectrum estimation.
This corresponds to the $k_{\perp}$ range $7 \times 10^{-3} \, {\rm Mpc}^{-1}$ to $2.03 \, {\rm Mpc}^{-1}$.
Figure  \ref{fig:1} shows  the binned power spectrum $\cl(\Delta \nu)$  at two different values of $\ell$ considering 
simulations with noise and $80 \%$ flagging.
For comparison we have also shown the theoretical 
prediction  corresponding to the input model power spectrum $P^m(k)$ calculated using 
\begin{equation} 
\cl(\Delta \nu)=\frac{1}{\pi r^2} \int_{0}^{\infty} d k_{\parallel} \, \cos(k_{\parallel} \, r^{'} \, \Delta \nu)
P(k_{\perp},k_{\parallel})
\label{eq:c1}
\end{equation}
which is the inverse of eq.~(\ref{eq:cl_Pk}). We see that  the 
results from the simulations are in agreement with the theoretical predictions. The results shown 
here are visually indistinguishable from the results from simulations with no noise and no flagging, 
or those with $20 \%, 40 \%$ and $60  \%$  flagging,  and we have not shown the other results here.  

We find that the value of $\cl(\Delta \nu)$
falls rapidly as $\Delta \nu$ is increased, and it has a value close to zero for $\Delta \nu > 1 \, {\rm 
MHz}$. Considering the simulations with noise and $80 \%$ flagging,  Figure \ref{fig:2} provides a visual representation of $\cl(\Delta \nu)$ across the entire $\ell$ and $\Delta \nu$ range that we have considered here. The results  are 
visually indistinguishable even if we have no noise and no flagging (or less flagging),  
or if we evaluate $\cl(\Delta \nu)$ analytically using (eq.~\ref{eq:c1}) and we have not shown these here. We see that the value of $\cl(\Delta \nu)$ decrease as  $\ell$ is increased. For a fixed $\ell$, the value of $\cl(\Delta \nu)$ falls rapidly as $\Delta \nu$ is increased and it has a value close to zero at large $\Delta \nu$.  The decrease in the value of $\cl(\Delta \nu)$ 
with increasing $\Delta \nu$ is more rapid as we go to larger $\ell$. The behaviour of $\cl(\Delta \nu)$ is directly 
manifested in the visibility correlation $V_2(U,\Delta \nu)=\langle \V^{*}(\U,\nu)\V(\U,\nu + \Delta \nu) \rangle$ with  $\ell=2 \pi U$. This  has been studied  extensively in several earlier works \citep{BS01,BP03,BA5}, and  we do not discuss this any further here.

\begin{figure*}
\begin{center}
\psfrag{kpara in Mpc-1}[1.0]{$\hspace{2cm}k_{\parallel}\hspace{.2cm} \textrm{Mpc}^{-1}$}
\psfrag{kper in Mpc-1}[1.0]{$\hspace{2.5cm}k_{\perp}\hspace{.2cm} \textrm{Mpc}^{-1}$}
\psfrag{K2 Mpc3}[c][1.0]{$P(k_{\perp},k_{\parallel})\hspace{.2cm} \textrm{mK}^{2} \textrm{Mpc}^{3}$}
\psfrag{All channels, avg. 24 realizations, 80
\psfrag{ 0.01}[c][1.0]{$ 0.01$}\psfrag{ 0.1}[c][1.0]{$ 0.1$}\psfrag{ 1}[1.0]{$ 1$}
\psfrag{10}{$ $}
\psfrag{-1}[1.0]{$\hspace{.2cm}10^{-1}$}\psfrag{-2}[1.0]{$\hspace{.2cm}10^{-2}$}\psfrag{0}[1.0]{$10^{0}$}\psfrag{1}[1.0]{$10^{1}$}\psfrag{2}[1.0]{$10^{2}$}\psfrag{3}[1.0]{$10^{3}$}
\includegraphics[scale=.8,angle=0]{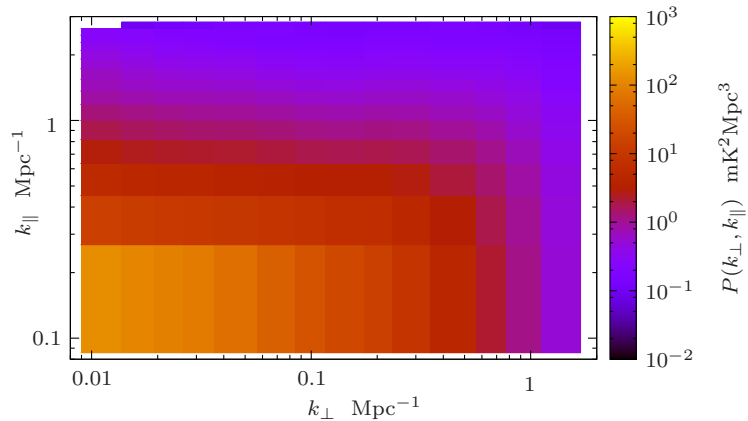}
\caption{The cylindrical power spectrum $P(k_{\perp},k_{\parallel})$ estimated from simulations with with noise and 80\% flagging.}
\label{fig:3}
\end{center}
\end{figure*}

Figure \ref{fig:3} shows the cylindrical power spectrum $P(k_{\perp},\,k_{\parallel})$  estimated 
by applying  eq.~(\ref{eq:b1})  to the $\cl(\Delta \nu)$ obtained from the simulations 
with  noise and $80 \%$  flagging (Figure \ref{fig:2}). We obtain estimates of $P(k_{\perp},\,k_{\parallel})$ in $15$  bins of equal logarithmic spacing along $k_{\perp}$  each with $N_c=257$ values along $k_{\parallel}$.  We have further binned $P(k_{\perp},\,k_{\parallel})$  into $16$  bins of equal logarithmic spacing along  $k_{\parallel}$  to increase the signal to noise ratio and also for convenience of plotting.
The results for the other cases which we have considered (lesser flagging, with/without noise) are very similar and they have not been shown separately. Note that the estimated power spectrum turns out to be negative at a single pixel (top left corner of the figure) when we have $80 \%$ flagging. In contrast, we obtain positive values at all the pixels when we consider a smaller percentage of flagged data. 

\begin{figure*}
\begin{center}
\psfrag{K2 Mpc3}[1.0]{$\hspace{2cm}P(k_{\perp},k_{\parallel})\hspace{.2cm} \textrm{mK}^{2} \textrm{Mpc}^{3}$}
\psfrag{All channels, avg. 24 realizations, 80
\psfrag{ 0.01}[l][1.0]{$ 0.01$}\psfrag{ 0.1}[l][1.0]{$ 0.1$}\psfrag{ 1}[1.0]{$ 1$}
\psfrag{10}{$ $}
\psfrag{-1}[1.0]{$\hspace{.2cm}10^{-1}$}\psfrag{-2}[1.0]{$\hspace{.2cm}10^{-2}$}\psfrag{0}[1.0]{$10^{0}$}\psfrag{1}[1.0]{$10^{1}$}\psfrag{2}[1.0]{$10^{2}$}\psfrag{3}[1.0]{$10^{3}$}
\psfrag{P(k)}[
1.0]{$P(k)\hspace{.2cm} \textrm{mK}^{2}\textrm{Mpc}^{3}$}
\psfrag{PM(k)-P(k)/PM(k)}[c][1.0]{$\delta$}
\psfrag{k}[1.0]{$k\hspace{.2cm}\textrm{Mpc}^{-1}$}
\psfrag{ 0.01}[1.0]{$ 0.01$}\psfrag{ 1}[1.0]{$ 1$}
\psfrag{-0.2}[1.0]{$-0.2$}\psfrag{-0.1}[1.0]{$ $}\psfrag{ 0}[1.0]{$ 0.0\hspace{0.4cm}$}\psfrag{ 0.1}[1.0]{$ $}\psfrag{ 0.2}[1.0]{$ 0.2$}
\psfrag{-0.15}{$ $}\psfrag{-0.05}{$ $}\psfrag{ 0.05}{$ $}\psfrag{ 0.15}{$ $}
\psfrag{None}[r][1.0]{$\textrm{No noise, No flagging} $}
\psfrag{n80}[r][1.0]{$\textrm{Noise, 80\% flagging} $}
\psfrag{model}[r][1.0]{$\textrm{Model}$}
%
\psfrag{-0.2}[1.0]{$-0.2\hspace{0.2cm}$}\psfrag{-0.1}[1.0]{$-0.1\hspace{0.2cm}$}\psfrag{ 0}[1.0]{$ 0.0\hspace{0.4cm}$}\psfrag{ 0.1}[1.0]{$0.1$}\psfrag{ 0.2}[1.0]{$0.2$}
\psfrag{-0.15}[1.0]{$ $}\psfrag{-0.05}{$ $}\psfrag{ 0.05}{$ $}\psfrag{ 0.15}[1.0]{$ $}
\psfrag{10}{$ $}
\psfrag{-1}[1.0]{$ $}\psfrag{0}[r][1.0]{$10^{0}$}\psfrag{1}[r][1.0]{$10^{1}$}\psfrag{2}[r][1.0]{$10^{2}$}\psfrag{3}[r][1.0]{$10^{3}$}\psfrag{4}[r][1.0]{$10^{4}$}\psfrag{5}[r][1.0]{$10^{5}$}
\includegraphics[scale=.8,angle=0]{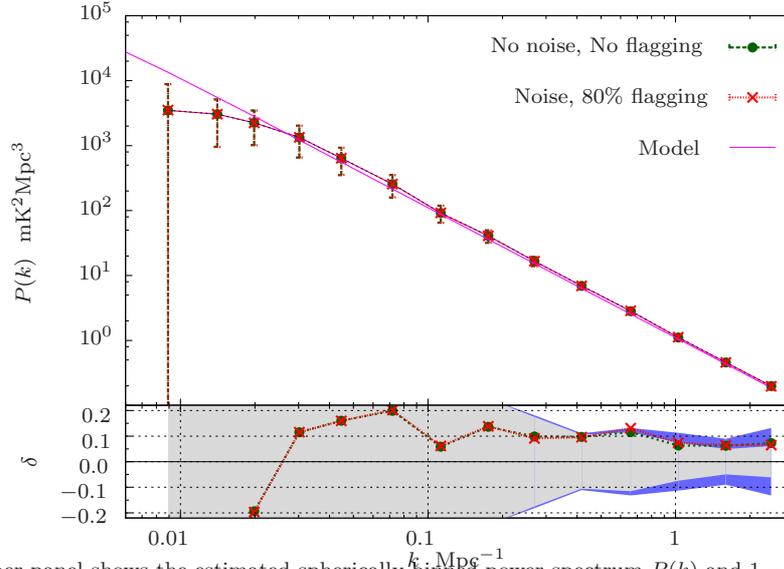}
\caption{The upper panel shows the estimated spherically-binned power spectrum $P(k)$ and $1-\sigma$ error-bars  for simulations with no noise and flagging and also with noise and $80 \%$ flagging. For comparison, the input model $P^{m}(k)$ is also shown by the solid line.
 The bottom panel shows the fractional error $\delta=[P(k)-P^m(k)]/P^m(k)$  (data points)  and the relative statistical fluctuation $\sigma/P^m(k)$ (shaded regions). The values of $\sigma$ are larger for simulations with noise and $80 \%$ flagging as compared to those with no noise and no flagging.}
\label{fig:4}
\end{center}
\end{figure*}

The upper panel of 
 Figure~\ref{fig:4} shows the spherical power spectrum $P(k)$ estimated from the simulations with no noise and no flagging,  and also the simulations with noise and $80 \%$ flagging. Here the 
 $P(k_{\perp},\,k_{\parallel})$ values  were combined  into $15$ bins of equal logarithmic interval in the $k$ range 
 $4 \times 10^{-3} \, {\rm Mpc}^{-1}$ to $3 \, {\rm Mpc}^{-1}$.
 We see that the results from these two sets of simulations are visually indistinguishable. The results for all the other cases considered here are very similar and they have not been shown separately. 
 The model power spectrum $P^{m}(k)$ is also shown for comparison. We see that $P(k)$ estimated from the simulations is below the model predictions at $k < 0.02 \, {\rm Mpc}^{-1}$.  Our estimator assumes that the convolution due to the telescope's primary beam pattern can be well approximated by a multiplicative factor 
which we have  incorporated in $M_g$  (eq.~\ref{eq:a4}).  Earlier studies \citep{samir14} show that this assumption does not hold at the small baselines (which also correspond to small $k_{\perp}$) that probe angular scales which are comparable to the angular extent of the telescope's primary beam pattern. 
The estimated power spectrum is in better agreement with the input model at $k \ge 0.02 \, {\rm Mpc}^{-1}$.  We however notice that $P(k)$ is somewhat  overestimated at $k \ge 0.03 \, {\rm Mpc}^{-1}$, but this difference goes down at larger $k$. 
The lower panel  shows the fractional deviation $\delta=[P(k)-P^m(k)]/P^m(k)$ of the estimated power spectrum $P(k)$ relative to the input model $P^m(k)$,  the shaded regions shows the $1-\sigma$ errors $\sigma/P^m(k)$.
We find that $P(k)$ is overestimated by $10 - 20 \%$ in the range  $0.03 \le k < 0.1 \, {\rm Mpc}^{-1}$, this falls to $5 - 15 \% $ in the range  $0.1 \le k < 1.0 \, {\rm Mpc}^{-1}$ and  the overestimate is 
less than $7.5 \%$ at $k > 1.0 \, {\rm Mpc}^{-1}$. There is around $\sim 1 \%$ difference in the estimated values when we have noise and $80 \%$ flagging as compared to the situation when these are not incorporated. Further, the values of $\sigma$ are larger when we introduce noise and flagging, this is particularly more pronounced at  large $k$.   In all cases we find that the errors  $\delta$ are less than the expected statistical fluctuations $\sigma/P^m(k)$.

\section{Discussion and conclusions} 
The error-bars shown here are based on $24$ independent realizations of the simulations. \citet{samir14} 
and \citet{samir16b} present analytical  formulas for estimating the statistical errors, and it is possible to obtain similar formulas for $\cl(\nu_a,\nu_b)$ and propagate the resulting errors through the Fourier transform to predict errors for $P(\k)$. However, simulations offer a more straight forward method to estimate the errors. It is possible to use the estimated power spectrum as an input for simulations, and use multiple realizations of these simulations to estimate the error-bars for the estimated power spectrum. 

An earlier study  \citep{samir14} used simulations to shows that the TGE overestimates $\cl$  due to the sparse and patchy  $uv$ coverage of the GMRT baseline distribution. This overestimate was found to come down if a more dense and uniform $uv$ coverage was  considered instead.  
The  estimator presented here overestimates $P(k)$ by $5-20 \%$ across a large portion of the $k$ range, the exact cause for this is not known at present .  We believe that this  is a consequence of the  sparse  and patchy  $uv$ coverage of the GMRT baseline distribution, and is not an inherent limitation of the estimator. We expect this effect to be much less severe for an array with a denser and more uniform $uv$ coverage. Further studies considering  arrays with different $uv$ coverage are needed to quantitatively establish this, and we propose to address this in future  work.

The signal in the visibility measurements $\V(\U_i,\nu_a)$ at different baselines $\U_i$ are not independent due to the telescope's primary beam pattern  and the signal at baselines within $D/\lambda$ 
are correlated (eq.~(12) of \citealt{BA5}) where $D$ is the antenna diameter.  Similarly, the 
visibility measurements $\V(\U_i,\nu_a)$  at different frequency channels $\nu_a$ are not independent (Figure 9 of \citealt{BA5}) and the signal remains correlated across different channels, the width of the correlation depending on the value of $\U_i$ \citep{BP03}. The signal contained in the  flagged data which is lost is also contained in the valid data which is available at our disposal for power spectrum estimation,  and the estimator presented here is able to recover the power spectrum equally well even if $80 \%$ of the data is flagged. While the estimated power spectrum is practically unchanged with or without flagging, the statistical fluctuations $\sigma$ are somewhat larger (particularly at large $k$) when flagging is introduced.   The 
entire analysis presented here is restricted to a situation where randomly chosen frequency channels were flagged. A variety of other situations may occur in  real life. For a given real data it would be best to first use the  flagging variables of the actual data in conjunction with simulations  to verify if the estimator can reproduce the input model of the simulation.  If needed, the discrete Fourier transform of eq.~(\ref{eq:b1}) 
can be replaced by a more sophisticated spectral estimator. However, here it is necessary  to apply this  to the final binned data and not the individual baselines, and therefore the problem is not computationally demanding. We propose to address these issues in more detail in future work. 

In a recent paper \citet{morales18} has  broadly classified the power spectrum estimators into two classes namely (1.) the delay spectrum or measured sky estimators, and (2.) the reconstructed sky estimators. The former class of estimators performs the Fourier transform from $\nu$ to $k_{\parallel}$ at a fixed antenna separation $\bf{d}$ which does not incorporate the frequency dependence of the baseline. In contrast, the same Fourier transform is carried out at the baseline $\u$ corresponding to a fixed angular scale which effectively incorporates the variation of baseline with frequency, however it uses a reconstructed sky model instead of the measured sky signal. The estimator presented here deals with the measured sky signal, it however differs from the usual delay spectrum estimators in that the signal is first correlated and then Fourier transformed. It is consequently possible to incorporate the frequency dependence of the baselines (as mentioned in Section  \ref{maps}).  This has not been incorporated in the present work, we plan to incorporate this and study its impact on foregrounds in future work.

\bibliographystyle{mnras}
\bibliography{refs}

\label{lastpage}

\end{document}